\documentclass{aptpub}

\usepackage{graphicx}
\date{\today}
\begin{document}
\title{Epidemic Size in the SIS Model of Endemic Infections}
\authorone[Bar-Ilan University]{David A. Kessler}
\addressone{Department of Physics, Bar-Ilan University, Ramat-Gan, IL52900 Israel}
\begin{abstract}
We study the Susceptible-Infected-Susceptible model of the spread of an endemic infection.  We calculate an exact expression for the mean number of transmissions for all values of the population and the infectivity.  We derive the large-$N$ asymptotic  behavior for the infectivitiy below, above, and in the critical region.  We obtain an analytical expression for the probability distribution of the number of transmissions, $n$, in the critical region.  We show that this distribution has a $n^{-3/2}$ singularity for small
$n$ and decays exponentially for large $n$.  The exponent decreases with the distance
from threshold, diverging to infinity far below and approaching zero far above.
\end{abstract}
\section{Overview}
The Susceptible-Infected-Susceptible (SIS) model of Weiss and Dishon~\cite{WeiDis} is one of the simplest model of endemic infections.  The model describes the evolution of an infection in a fixed population, with no restriction on the possibility of reinfection of a previously infected and now recovered individual.  This is contrast to the venerable
Susceptible-Infection-Recovered (SIR) model~\cite{McKen}, where reinfection is not permitted.   Both
models exhibit a threshold value of the infectivity, below which the infection immediately
dies out.  Below threshold, then, where only a tiny fraction of the population is impacted, the two models have essentially equivalent statistical properties. Above threshold, in the SIR model the infection is self-limiting, since in a fixed population the number of potential new victims, the susceptible pool, is monotonically decreasing in size. The SIS model, on the other hand, describes  an endemic infection which can (above threshold) persist indefinitely, at least at the deterministic level.  Thus, the statistics of infection size in the two models above threshold are very different.   

The statistics of the mean time to extinction in the SIS model have been much studied, starting with the original paper of Weiss and Dishon~\cite{WeiDis} and most recently by Doering, Sargsyan and Sander~\cite{SanDoer}.  The latter paper investigates the large population
limit of the mean extinction time.  This goes from a logarithmic dependence on the population size, $N$, below threshold, to a $\sqrt{N}$ dependence exactly at threshold
to an exponential dependence above.  In this paper we will focus on the mean {\emph{number}} of transmissions till extinction. This is a more pertinent method of characterizing the epidemic and the threshold transition. We shall derive an exact formula for this
quantity for general $N$ and infectivity, and then examine its large $N$ asymptotics. As we shall see, above threshold the mean epidemic size is directly
related to the mean epidemic duration.  At and below threshold, though, these quantities
are quite different.  Furthermore, the number of infection events is directly relevant when 
considering the probability of a mutation of the pathogen, as mutations are most probable during the exponential growth phase following a new infection\cite{Antia}.  These mutations are implicated in the conversion of a sub-threshold weakly transmittable
pathogen into a super-threshold variety capable of inducing a major epidemic.

Of particular interest will be the critical regime separating the above and below threshold
cases.  As already noted by Nasell~\cite{Nasell}, for a range of infectivities of width $1/\sqrt{N}$ around threshold, there is a crossover region that interpolates between the above and below threshold cases.  The existence of a large $N$ scaling theory in this
region was recently proven by Dolgoarshinnykh and Lalley~\cite{DolgLal}.  We shall see
this crossover region and its characteristic scaling arising naturally from our general result for the mean infection size.  

After this treatment, dealing exclusively with the perhaps most biologically relevant case of a single initial infection, we extend our results to an arbitrary number of initial infections, again deriving an exact formula and then examining the large $N$ asymptotics.  In the crossover regime, we will have to distinguish the cases when the number of initial infections in small, comparable to, or much larger than $\sqrt{N}$.

From looking only at the mean number of infections, we move on to consider the entire probability distribution for the number of infections.  We first briefly discuss the above and below threshold cases, and then focus in on the critical threshold regime.  In the particular case of exactly at threshold, the entire probability distribution for the appropriate scaling variable (the number of infections divided by $N$) can be explicitly displayed.  In general, we can express the probability distribution as an inverse Laplace transform.  This is sufficient to calculate the limiting behavior of the distribution for small and large epidemics, and to recover our expression for the mean in the critical regime.
We then conclude with a few observations.

\section{Preliminaries}

We begin with a description of the SIS model.  The $N$ individuals in the population are divided into two subclasses: the susceptible pool, of size $S$, and the infected (and infectious) class, of size $I$, with $N=S+I$. The disease is transmitted from an infected individual to a susceptible one with rate $\alpha/N$, so that
$$
(S,I)\stackrel{\alpha S I/N}{\to} (S-1,I+1) .
$$
Infected individuals recover with a rate $\beta$, reverting back to susceptibles:
$$
(S,I)\stackrel{\beta I}{\to} (S+1,I-1).
$$
Of primary interest is the case where initially $S=N-1$, $I=1$, so that the outbreak is sparked by a single infected individual.  The outbreak  terminates when the last infected individual recovers, and $I$ returns to 0.

This stochastic process is traditionally approximated (for large populations) by the rate equations
\begin{eqnarray*}
\dot{S}&=& - \frac{\alpha}{N} S I + \beta I\nonumber\\
\dot{I}&=& \frac{\alpha}{N}SI - \beta I 
\end{eqnarray*}
Using the conservation of $N$, we get
$$
\dot{I} = (\alpha - \beta) I -\frac{\alpha}{N}I^2 
$$
which is a logistic-type equation.  We see that there is a transitition at 
 $R_0 \equiv \alpha/\beta=1$, where $R_0$ is equal to the mean number of primary infections caused in a large population of susceptibles by an infected individual.
It is clear that if ${R_0} <  1$, the $I=0$ state is stable, whereas for $R_0 > 1$ the
 rate equation predicts  a stable equilibrium state at $S_* =( \beta/\alpha )N$,
$I_* = (1 - \beta/\alpha)N$.
Thus at the classical level, ${R_0}=1$ marks the threshold between an infection that becomes endemic and those that fail to spread.

\section{Mean Number of Infections}

Already in the original Weiss-Dishon paper~\cite{WeiDis}, an exact expression for the mean time to extinction, starting from the completely infected state, was derived.  The generalization
of this to an arbitrary number of initial infected individual was given in Leigh~\cite{Leigh}
and rediscovered by Doering, et al. ~\cite{SanDoer}.  However, the mean {\emph{number}} of infections is the quantity of primary interest in an infection model.  We
can focus in on this quantity if we  eliminate time, considering only the transitions between states. We characterize the system by the
number, $T$, of transitions the system has undergone. In each transition the number of infected individuals either rises or falls by one, so that $I$ undergoes a kind of random walk. The probability of an upward transition is $p^+={R_0} S/({R_0} S + N)=R_0 (N-I)/(R_0(N-I) + N)$, whereas the probability of a downward transition is $p^-=1-p^+$.  These probabilities
are unequal and depend on $I$, so that the walk is biased, with a  "space" -dependent drift.  (From here on, we will  refer to $T$ as Time, with the lower case
word "time" retaining its usual meaning, and trust this will not lead to confusion). It is easy to see that at the point of extinction,  the total number of infections, including the initial $n_o$ infected individuals, is just $n=(T_{\textit{ext}} + n_o)/2$.  The number of induced infections is of course $n_o$ smaller.  Since
the results of Weiss and Dishon, Leigh and Sander, et al. for the mean time to extinction apply to
a general one-step random walk, we can apply them directly to calculate the mean number of infections. Specializing to the case where we initially have exactly one infected, we have for the mean extinction Time, $\tau_1$
$$
\tau_1 = \sum_{j=1}^{N} \frac{1}{p^-_{j}}\prod_{k=1}^{j-1} \frac{p^+_k}{p^-_k}
$$
Here we have indicated explicitly the dependence of the transition probabilities on $I$
via a subscript.  Plugging in these probabilities,
we find
$$
\tau_1 = \sum_{j=1}^{N}  \left(1 + R_0 - \frac{R_0 j}{N}\right)\frac{R_0^{j-1} (N-1)!}{N^{j-1}(N-j)!}
$$
Reordering the sum, we can rewrite this as
$$
\tau_1 = \left(\frac{R_0}{N}\right)^{N-1}(N-1)!\sum_{n=0}^{N-1}\left(1 +  \frac{R_0 n}{N}\right) \frac{1}{n!}\left(\frac{N}{R_0}\right)^n
$$
We can do better, since the second term in the sum in the same as the first, except for
the last index, so
\begin{eqnarray*}
\tau_1&=&\left(\frac{R_0}{N}\right)^{N-1}(N-1)!\left[2\sum_{n=0}^{N-1} \frac{1}{n!}\left(\frac{N}{R_0}\right)^n \right.\nonumber\\
&\ &\quad\quad\quad\quad\quad\quad\quad\quad\quad\quad \left. {} - \frac{1}{(N-1)!}\left(\frac{N}{R_0}\right)^{N-1}\right]\nonumber\\
&=&2 \left(\frac{R_0}{N}\right)^{N-1}(N-1)!\sum_{n=0}^{N-1} \frac{1}{n!}\left(\frac{N}{R_0}\right)^n - 1
\end{eqnarray*}
so that
$$
\bar{n}_1 = \left(\frac{R_0}{N}\right)^{N-1}(N-1)!\sum_{n=0}^{N-1} \frac{1}{n!}\left(\frac{N}{R_0}\right)^n
$$
We recognize the sum as the first $N$ terms of the Taylor expansion of the exponential
$e^{N/R_0}$. The behavior of the sum depends on whether $R_0$ is above or below 1.
This follows from the fact that the terms in the Taylor expansion of $e^x$ increase until
$n=x$, and then decrease.  For large $x$, in fact, the behavior of the terms with $n$ is a Gaussian peaked at $n=x$.  The behavior of the sum is then determined by whether
 the last term of the sum at $n=N-1$ occurs before or
after the peak at $n=N/R_0$, i.e. whether $R_0$ is above or below 1.

\begin{figure}
\center{\includegraphics[width=0.7\textwidth]{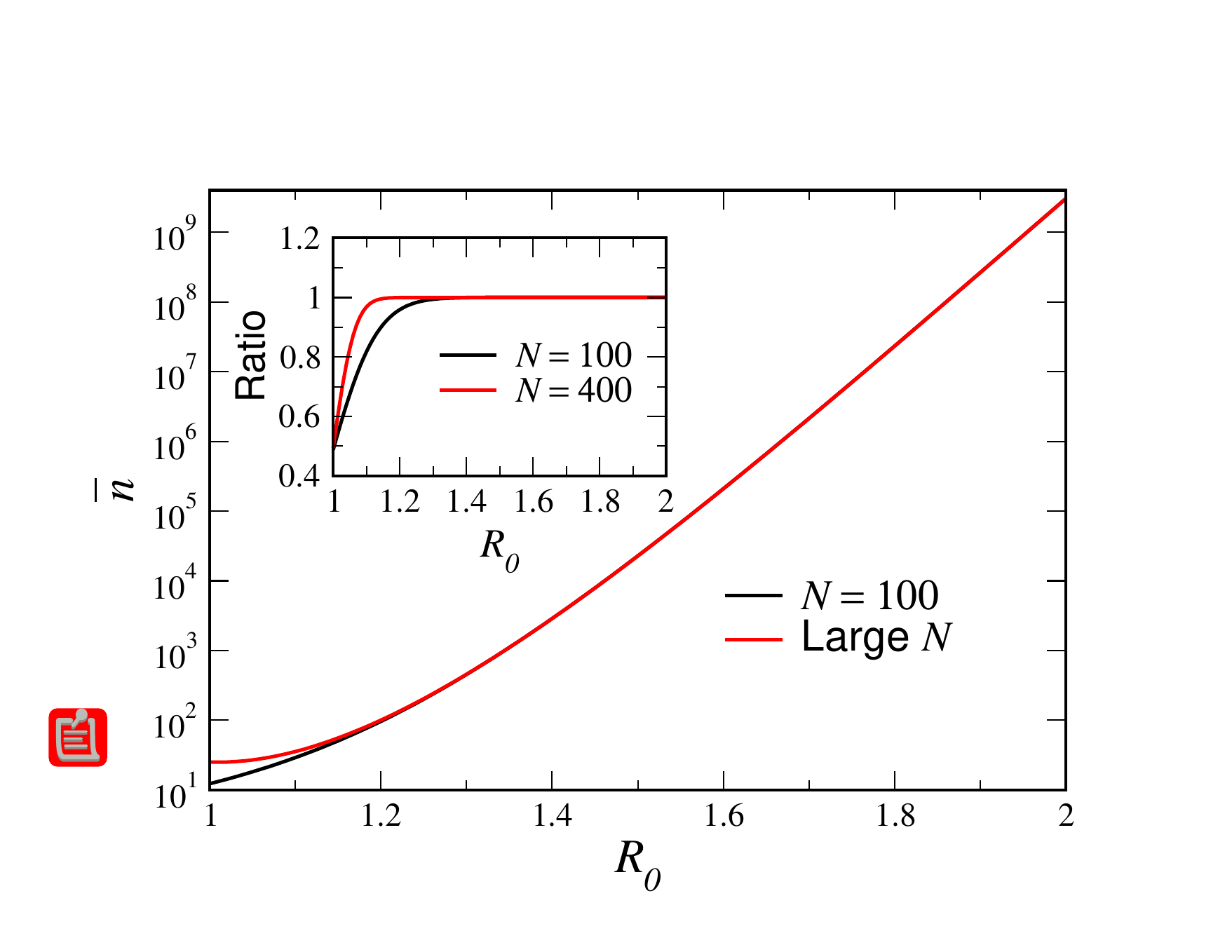}}
\caption{Supercritical Regime ($R_0>1$): Mean number of infections, $\bar{n}$ versus the infectivity parameter $R_0$,  for the case of starting for one infected individual.  Also shown is  the 
large $N$ asymptotic formula, Eq. (\ref{super}).  The cases of population $N=100$ is shown. Inset: Ratio of the exact mean number of infections, $\bar{n}$ to that given by the
large $N$ asymptotic formula, Eq. (\ref{super}), versus the infectivity parameter $R_0$. The cases of population $N=100, 400$ are shown.}
\label{sup100}
\end{figure}

For $R_0$ above 1, the summed terms extend past the peak, which dominates the sum, and so,
up to exponentially small corrections, the sum is just the exponential.  Furthermore the
prefactor can be approximated via Stirling's formula, giving
\begin{equation}
\bar{n}_1 \approx\frac{ \sqrt{2\pi N}}{R_0} e^{N(\ln(R_0) + 1/R_0 -1)}
\label{super}
\end{equation}
Thus, as expected the mean number of infected cases grows exponentially large with
$N$, with the exponent going to 0 as $R_0$ approaches 1.  Furthermore, the exponent is the same as for the mean first passage time (here actual time) as calculated in Ref. 
\cite{SanDoer}, and is equal to the action for the semiclassical extinction trajectory~\cite{Kamenev,KessWKB}.  This is because above threshold, the system remains an exponentially long time in the classically stable state. We plot $\bar{n}$ versus $R_0$
in Fig. \ref{sup100}, together with the large-$N$ asymptotic formula, Eq. (\ref{super}).
We see that the mean number of infections quickly grows to astronomical proportions as
$R_0$ increases away from 1.  To see the approach to the large-$N$ result, we show in the inset the ratio of the exact results for
$N=100$ and $400$ to the large-$N$ asymptotic formula.  We see that the approximate formula works excellently except in the
vicinity of the transition point $R_0 = 1$, and improves with increasing $N$.

\begin{figure}
\center{\includegraphics[width=0.7\textwidth]{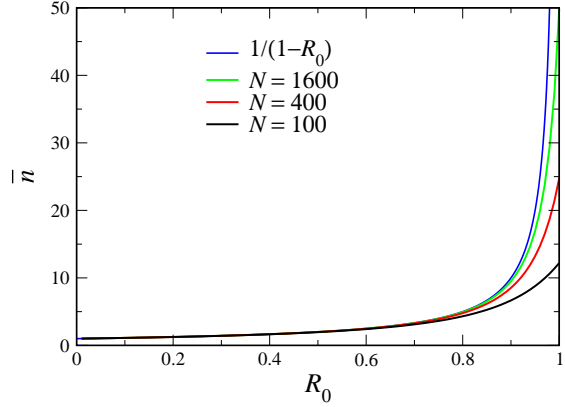}}
\caption{Subcritical Regime ($R_0 < 1$): Mean number of infections, $\bar{n}$, versus the infectivity parameter $R_0$, when starting with one infected individual, together with the prediction for an infinite population.  The cases of population $N=100, 400, 1600$ are shown.}
\label{sub}
\end{figure}

For $R_0$ below 1, the sum is cut off while the terms are still increasing with $n$, and so the largest terms in the sum are the last ones, which approximate
a geometric series
$$
\bar{n}_1 \approx \sum_{k=0}^\infty R_0^k = \frac{1}{1-R_0}
$$
which of course is the same as in the SIR model, since the number of infected persons
is so small that no one gets a multiple infection.  This infinite $N$ answer is compared to the finite $N$ results in Fig. \ref{sub}, where we see that it works well as long as we
are sufficiently below $R_0=1$, and the range of agreement increases with $N$.  As opposed to the supercritical case, here there is in general no simple relation between the
mean number of infections and the mean (actual) time to extinction, which is given by
$$
\bar{t}_1 = \sum_{k=1}^N \frac{1}{\beta k} \left(\frac{R_0}{N}\right)^{k-1} \frac{(N-1)!}{(N-j)!}
\approx \sum_{k=1}^\infty \frac{1}{\beta j} R_0^{k-1} = -\frac{1}{\beta R_0} \ln(1-R_0)
$$
This is of course due to the fact that in the subcritical  case the number of infections in not strongly peaked about some value as it was in the supercritical case.

\begin{figure}
\center{\includegraphics[width=0.7\textwidth]{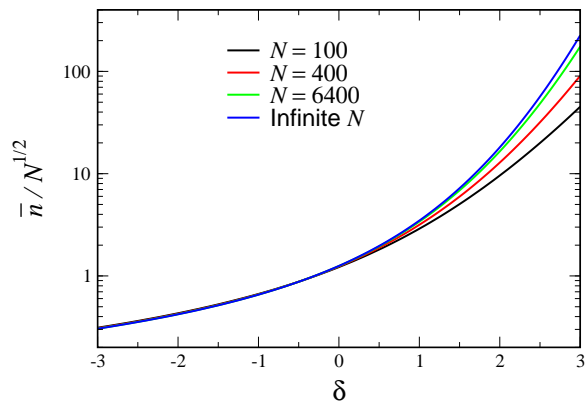}}
\caption{Critical Regime ($R_0 = 1 + \delta/\sqrt{N}$): Scaled mean number of infections,
$\bar{n}/\sqrt{N}$ versus scaled infectivity parameter $\delta$, when starting with one
infected individual.  Also shown is the large $N$ asymptotic result, Eq. (\ref{crit}). The cases of population $N=100, 400, 6400$ are shown.}
\label{critfig}
\end{figure}

For $R_0$ near 1, $R_0 = 1 + \delta N^{-1/2}$, there is a transition region.  The
dominant terms in the sum are again the largest, which have a Gaussian character,
with a maximum at $N/R_0 \approx N$. 
\begin{eqnarray}
\bar{n}_1 &=& \sum_{k=0}^{N-1} R_0^k\prod_{j=0}^{k-1} \left(1-\frac{j}{N}\right) \nonumber\\
&\approx& \sum_{k=0}^{N-1}\exp\left(k\left(\frac{\delta}{\sqrt{N}} \right)-\frac{k^2}{2N}\right)\nonumber\\
&\approx&\int_0^\infty dk \, \exp\left(k\left(\frac{\delta}{\sqrt{N}} \right)-\frac{k^2}{2N}\right)\nonumber\\
&=&\sqrt{\frac{\pi N}{2}}e^{\delta^2/2}\left[1 + \textrm{erf}\left(\frac{\delta\sqrt{2}}{2}\right)\right]
\label{crit}
\end{eqnarray}
This clearly reproduces the sub- and supercritical results in the limit of large negative and
positive $\delta$, respectively.
This formula is plotted in Fig. \ref{critfig} along with data for $N=100, 400, 6400$.  We see that finite $N$ data converge to the infinite $N$ prediction, with the finite $N$ effects
larger at larger $\delta$.

\section{Mean Number of Infections, General Initial Condition}

\begin{figure}
\center{\includegraphics[width=0.7\textwidth]{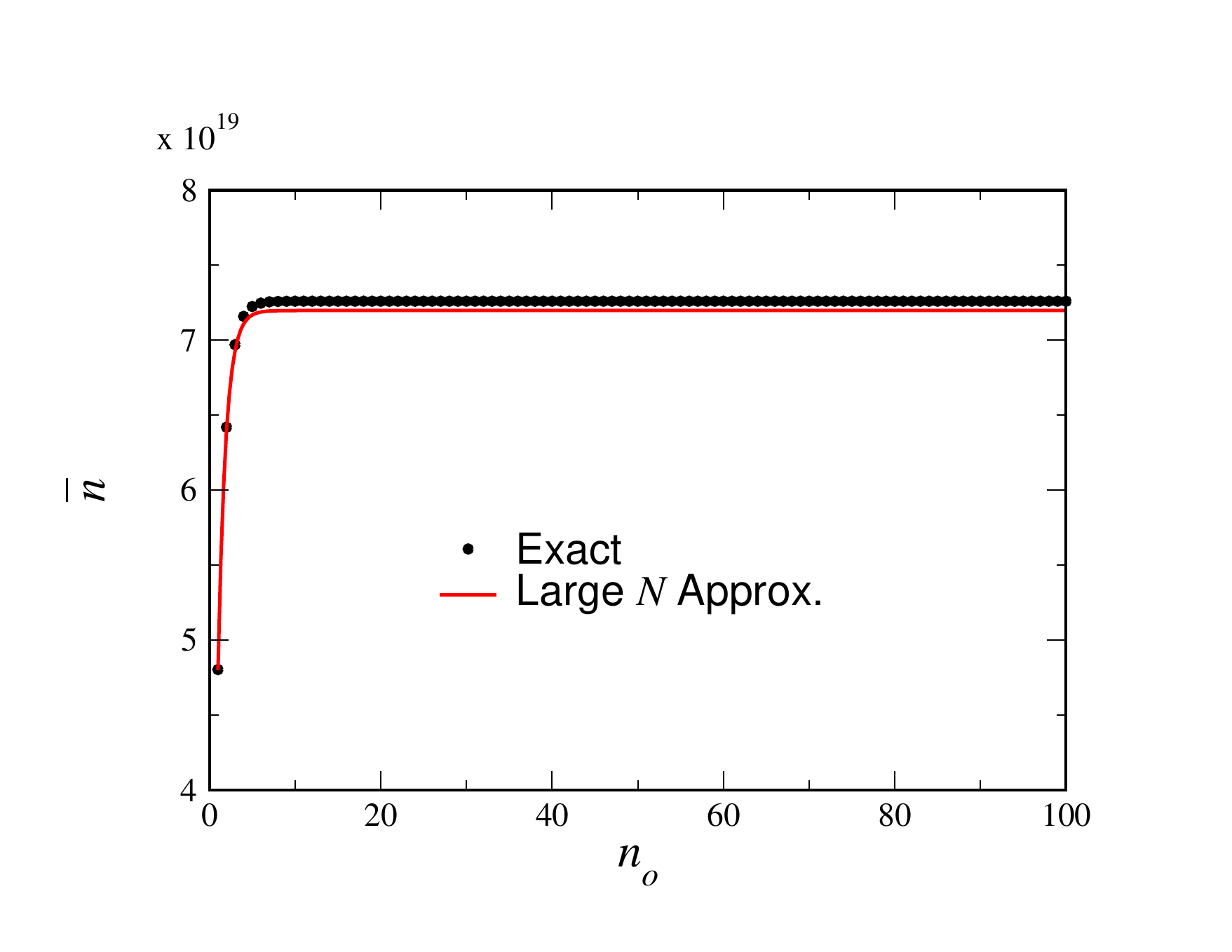}}
\caption{Exact calculation of $\bar{n}$ as a function of $n_o$ for the supercritical case, $R_0=3$ together with the analytical approximation, Eq. (\ref{nbar_over}).   $N=100$.}
\label{nbar100_3}
\end{figure}

These results are easily generalized to the case of $n_o$ initial infected individuals,
again starting from the corresponding mean first passage Time.  We get
$$
\bar{n}_{n_o} = \bar{n}_1 + \sum_{k=2}^{n_o} \sigma_k
$$
where
$$
\sigma_k \equiv \left(\frac{R_0}{N}\right)^{N-k} (N-k)! \sum_{j=0}^{N-k} \frac{1}{j!} \left(\frac{N}{R_0}\right)^j
$$
Again these results are instructive in the various limits.  For the above threshold case,
$\sigma_k \approx R_0^{1-k} \bar{n}_1$ so that
\begin{eqnarray}
\bar{n}_{n_o} &\approx& \left(1 + \frac{1}{R_0} + \frac{1}{R_0^2} + \ldots + \frac{1}{R_0^{n_o-1}}\right) \bar{n}_1 \nonumber\\
&\approx& \left(1 - R_0^{-n_o}\right) \left[ \frac{\sqrt{2\pi N}}{R_0 - 1} e^{N(\ln(R_0) + 1/R_0 -1)}\right]
\label{nbar_over}
\end{eqnarray}
The prefactor is recognized as the probability of a biased random walk starting at
$n_o$ to survive to infinity.  Thus, the mean number of infections is the mean number 
of infections starting in the macroscopically infected metastable state times the probability of surviving long enough to reach this state.  The expression in brackets,
the mean number of infections starting in the metastable state, $\hat{n}_{(R_0-1)/R_0}$
is itself simply related to the mean first passage time for this initial state calculated
in Ref. \cite{SanDoer}. If one accounts for the average time for a transition
in the metastable state, $2(R_0-1)/NR_0$, one can easily obtain from the above the average Time
to extinction, which is twice the average number of infections.  This is because the overwhelming majority of the time is spent in the vicinity of the metastable state.
In Fig. \ref{nbar100_3}, we
present the exact results for $\bar{n}_{n_o}$ for the case $N=100$, $R_0=3$ together with our approximation, Eq. (\ref{nbar_over}).  We see
the agreement is quite satisfactory.

\begin{figure}
\center{\includegraphics[width=0.7\textwidth]{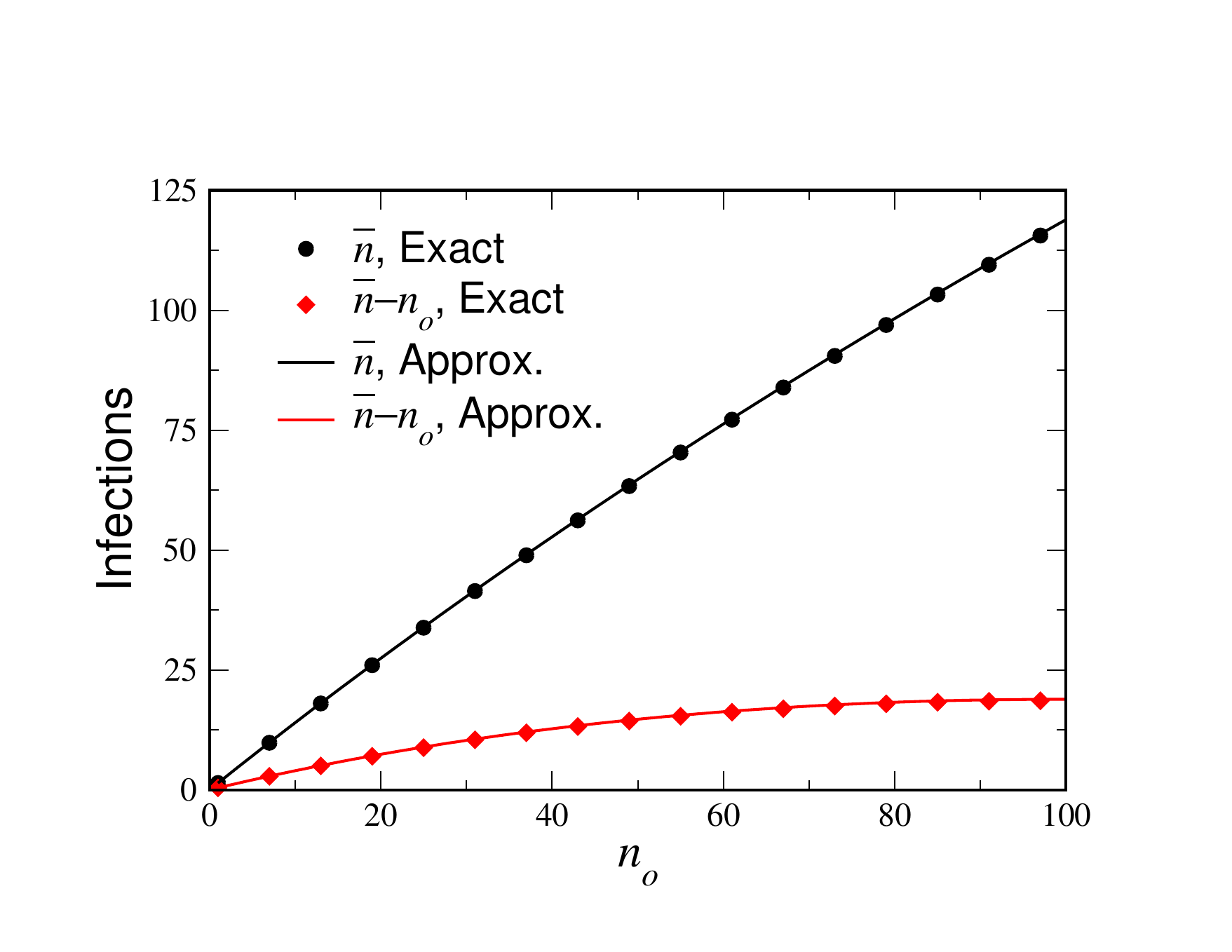}}
\caption{Exact calculation of $\bar{n}$ as a function of $n_o$ for the subcritical case, $R_0=0.3$ together with the analytical approximation, Eq. (\ref{nbar_under}).   Also
shown is the average number of {\emph{induced}} infections, $\bar{n}_{n_o}-n_o$.$N=100$.}
\label{nbar100_p3}
\end{figure}

Below threshold, 
$$
\sigma_k \approx  \sum_{j=0}^\infty\left[ \left(1-\frac{k}{N}\right)R_0\right]^j = \frac{1}{1-(1-k/N)R_0}
$$
Thus, for $n_o \ll N$ all the $\sigma$'s are all approximately equal to $\bar{n}_1$,  so that $
\bar{n}_{n_o} \approx n_o \bar{n}_1$. This is clear, as the individual seeds act essentially independently, since they impact an
infinitesimal fraction of the total population.  This is in sharp contrast to the above threshold case, where $\bar{n}_{n_o}$ converges exponentially quickly over an ${\cal{O}}(1)$ range of $n_o$.
For larger $n_o$ of order $N$, we get
\begin{equation}
\bar{n}_{n_o} \approx \int_0^{n_o} \sigma_k dk = \frac{N}{R_0} \ln\left(\frac{1-(1-n_o/N)R_0}{1-R_0}\right)
\label{nbar_under}
\end{equation}
For the subcritical case, it is also interesting to consider the number of \emph{induced}
infections, since here most of the infections are just those of the initial state.  For small
$n_o$ the number of induced infections is approximately $n_o R_0/(1-R_0)$, again
proportional to $n_o$.  For $n_o=N$, the number of induced infections is $(N/R_0) \ln(1/(1-R_0))-N$, which for small $R_0$ is approximately $NR_0/2$.  Thus the interference
between different initial seeds reduces the number of induced infections roughly by half in this case.  The interference effect is of course even more dramatic for larger $R_0$.  As $R_0$ approaches unity, the number of induced infections diverges only logarithmically for $n_o=N$, as opposed to the $1/(1-R_0)$ divergence for small $n_o$.
Of course, for $R_0$ even larger, in the supercritical regime, as we have seen, the interference effect is almost total, as increasing $n_o$ beyond 10 or so has essentially
no effect. The subcritical case is demonstrated in Fig. \ref{nbar100_p3} for the case $R_0=0.3$.

\begin{figure}
\center{\includegraphics[width=0.7\textwidth]{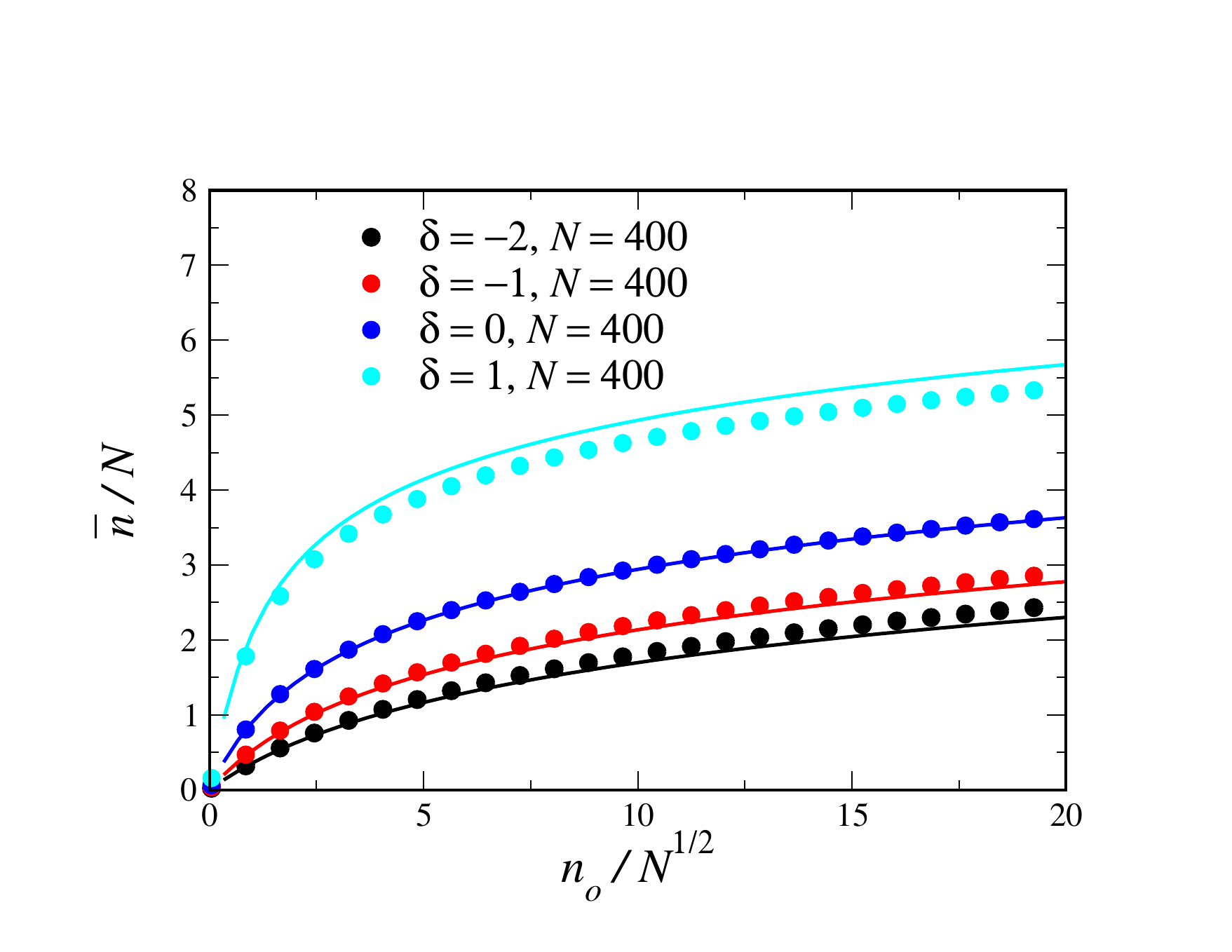}}
\caption{Scaled average infection size $\bar{n}/N$ as a function of the scaled initial number of
infected individuals, $n_o/\sqrt{N}$ for $N=400$ and $\delta = -2, -1, 0, 1$ ($R_0 = 0.9, 0.95, 1, 1.05$ respectively).}
\label{xctvscrit}
\end{figure}

In the critical regime, things are of course a bit more complicated.  Again, we first compute the $\sigma_k$:
\begin{eqnarray*}
\bar{\sigma}_k &=& \sum_{j=0}^{N-k} R_0^j\prod_{\ell=0}^{j-1} \left(1-\frac{k+j}{N}\right) \nonumber\\
&\approx& \sum_{j=0}^{N-k}\exp\left(j\left(\frac{\delta}{\sqrt{N}}-\frac{k}{N} \right)-\frac{j^2}{2N}\right)\nonumber\\
&=&\sqrt{\frac{\pi N}{2}}e^{(\delta-k/\sqrt{N})^2/2}\left[1 + \textrm{erf}\left(\frac{\sqrt{2}}{2}\left(\delta - \frac{k}{\sqrt{N}}\right)\right)\right]
\end{eqnarray*}
We now have to integrate this with respect to $k$, whose typical scale is ${\cal{O}}(\sqrt{N})$:
\begin{equation}
\bar{n}_{n_o} =N \sqrt{\frac{\pi }{2}} \int_0^{n_o/\sqrt{N}} dx\, e^{(\delta-x)^2/2}\left[1 + \textrm{erf}\left(\frac{\sqrt{2}}{2}\left(\delta - x\right)\right)\right]
\label{critnbarno}
\end{equation}
One immediate result is that $\bar{n}_{n_o}$ starts out as ${\cal{O}}(\sqrt{N})$ for 
small $n_o$ of order unity, but for $n_o$ of order ${\cal{O}}(\sqrt{N})$, the average
number of infections rises to ${\cal{O}}(N)$.  In Fig. \ref{xctvscrit}, we present the
exact results for $\bar{n}_{n_o}$ in the critical region for $N=400$ versus our scaling
prediction, Eq. (\ref{critnbarno}).  We see that the scaling results are perfect for the exactly
critical case, and the further we are from criticality, the larger the finite $N$ effects are.
Furthermore, the finite $N$ effects are larger for positive $\delta$ than for negative.
Also, the larger the initial infection size, the larger the finite $N$ effects.

\begin{figure}
\center{\includegraphics[width=0.7\textwidth]{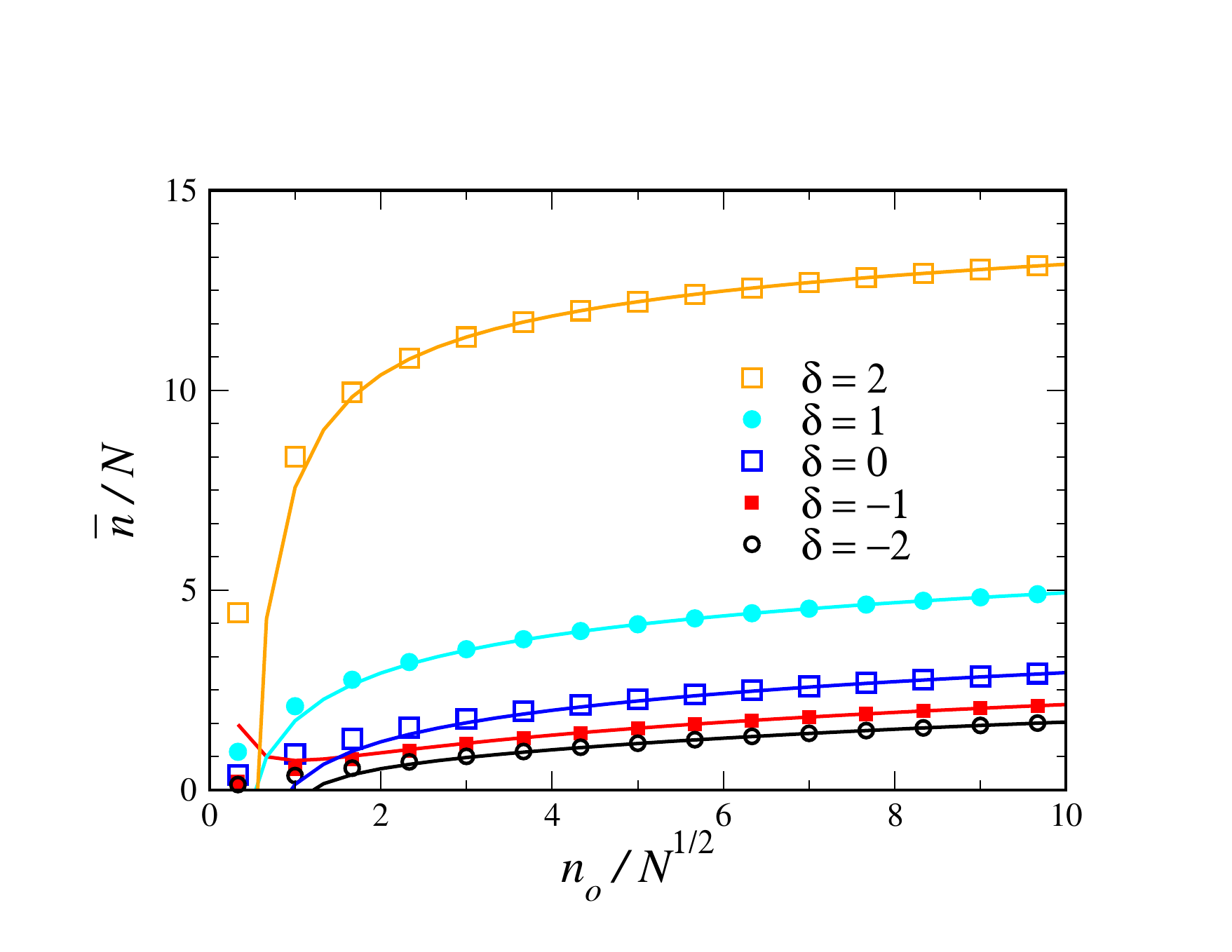}}
\caption{Calculation of scaling form of scaled mean epidemic size $\bar{n}/N$ as a function of scaled initial epidemic size $n_o/\sqrt{N}$ in the critical regime for intermediate $\delta=-2, -1, 0, 1, 2$.  Also plotted is the large $n_o$ analytic approximation,  Eq. (\ref{largeno}). }
 \label{many_del}
\end{figure}

\begin{figure}
\center{\includegraphics[width=0.7\textwidth]{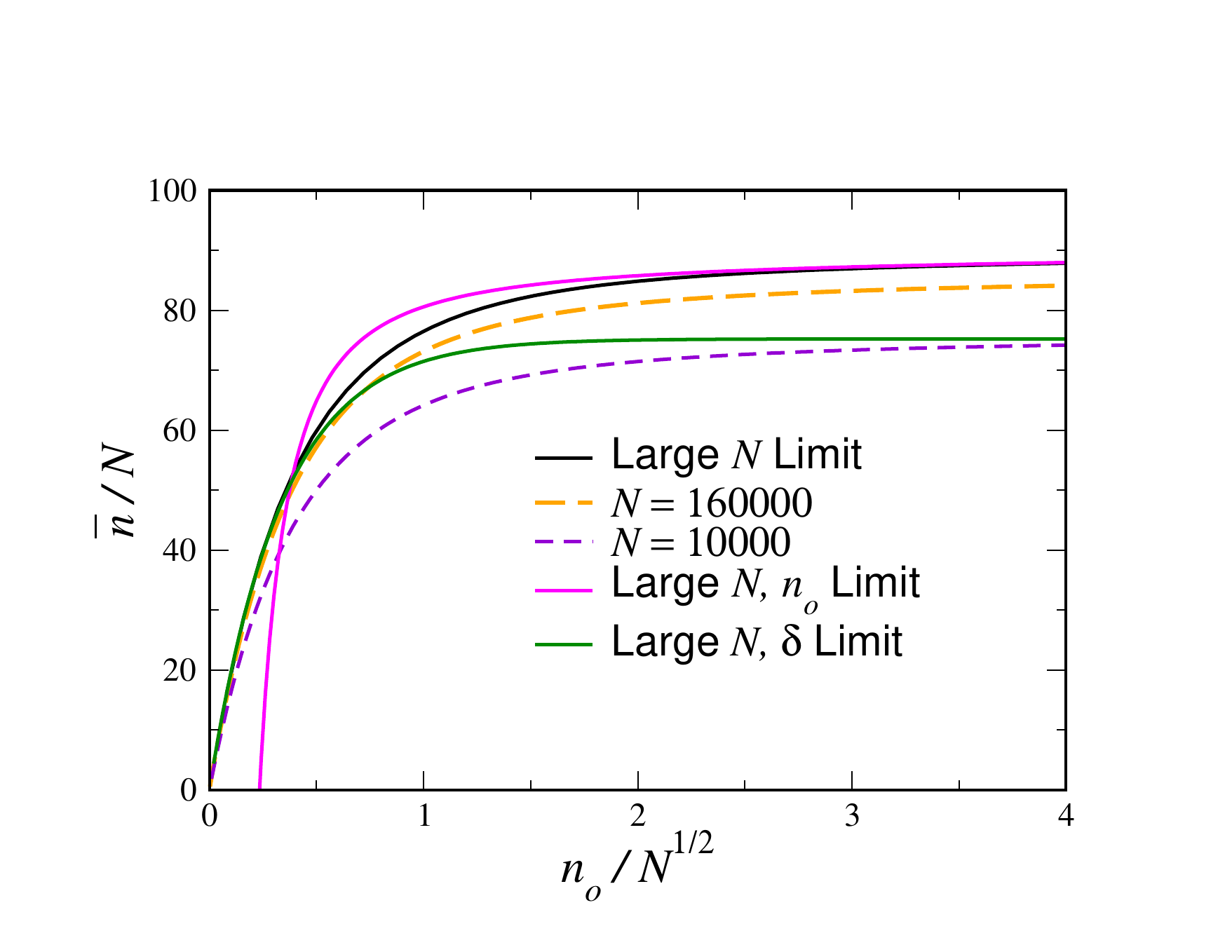}}
\caption{Scaling average epidemic size $\bar{n}/N$ as a function of scaled initial epidemic size $n_o$ for $N=10^4, 1.6\cdot10^5$ for the above critical case $\delta=3$. Also shown in the analytic critical scaling form, Eq. (\ref{critnbarno}), and its large $n_o$ (Eq. (\ref{largeno})) and large $\delta$ (Eq. (\ref{bigdel})) limits. }
 \label{largeD}
\end{figure}

To better understand our scaling formula, we consider in turn the cases $\delta$ large and negative,
$\delta$ of order unity, and $\delta$ large and positive.  In the former case, we the argument of the
{\emph{erf}} is large and negative and so
$$
\bar{n}_{n_o} \approx \sqrt{N}\int_0^{n_o} dk \frac{1}{|\delta| + k/\sqrt{N}} = N\ln\left(1+\frac{n_o}{|\delta|\sqrt{N}}\right)
$$
Thus, $\bar{n}_{n_o}$ crosses over from a linear behavior at $n_o$ of order unity, to 
logarithmic growth when $n_o$ of order $\sqrt{N}$.
For $\delta=0$ of order unity, it is more useful to integrate first with respect to $k$ and then
do the integral over $j$.  This gives the  formula
$$
\bar{n}_{n_o} = N\int_0^\infty dj e^{-j^2/2N}e^{\delta j} \frac{1-e^{-jn_o/N}}{j}
$$
Again, for $n_o$ small compared to $\sqrt{N}$, the answer is proportional to $n_o$.  For large $n_o$, this
can be approximated as follows:
\begin{eqnarray}
\bar{n}_{n_o} &\approx& \lim_{\epsilon\to 0^+} N\left[\int_0^\infty dj e^{-j^2/2N} j^{\epsilon-1} e^{j\delta}
- \int_0^\infty dj e^{-jn_o/N}  j^{\epsilon-1} (1 + j\delta + \ldots)(1 - \frac{j^2}{2} + \ldots)\right]\nonumber\\
&\approx& N\left[\frac{1}{2}\gamma + \ln(n_o\sqrt{2/N}) + A(\delta) - \frac{\delta\sqrt{N}}{n_o} - \frac{N(\delta^2 - 1)}{2n_o^2}\right]
\label{largeno}
\end{eqnarray}
where $A(\delta)$ is given be
$$
A(\delta) \equiv \sqrt{\frac{\pi}{2}} \int_0^\delta e^{t^2/2} dt + \int_0^\infty \frac{\cosh k\delta - 1}{k} e^{-k^2/2} dk 
$$
$A$ is a monotonically increasing function of $\delta$, with $A(0)=0$.  For large negative
$\delta$, $A(\delta) \approx -\ln(\delta) - \gamma/2 - (\ln 2)/2$, reproducing
our previous result.  For large positive
$\delta$, $A$ grows quickly, $A(\delta) \approx \sqrt{2\pi} e^{\delta^2/2}$.  Thus,
for all $\delta$, $\bar{n}$ grows logarithmically in $n_o$ for $n_o/\sqrt{N}$ sufficiently
large.  However, since $A$ grows so rapidly with $\delta$, for large positive $\delta$
this behavior is not readily visible in practice, since $n_o$ can be no bigger than $N$.
In Fig. \ref{many_del}, we show $\bar{n}_{n_o}$ for different $\delta$'s.  We see
that the large $n_o$ approximation works well for $n_o/\sqrt{N}$ larger than 1 or so.
For large positive $\delta$,  the argument of the \emph{erf} is large and positive, and so the behavior for fixed $n_o/\sqrt{N}$ is most relevant.  Then,
\begin{equation}
\bar{n}_{n_o} \approx \sqrt{2\pi N}e^{\delta^2/2} \int_0^{n_o} dk e^{-\delta k/\sqrt{N}} 
= N\sqrt{2\pi} e^{\delta^2/2} (1 - e^{-\delta n_0/\sqrt{N}})/\delta
\label{bigdel}
\end{equation}
The problem with this expression is that it is not at all accurate until $\delta$ is fairly
large, around 6 or so.  For such large $\delta$'s, the concept of a critical region does
not apply until really large $N$'s.  For example the next order correction to the action is
of order $\delta^3/N^{1/2}$, which is only small for $N \sim \delta^6$.  We can see this
in Fig. \ref{largeD}, where we examine the convergence of the finite-$N$ results to the
critical scaling result for $\delta=3$.  Only for $N=160,00$ is $\bar{n}$ approaching its
limiting scaling form.  This form is well approximated by our large $n_o$ formula, Eq. (\ref{largeno}) for $n_o/\sqrt{N}>2$, and by the large $\delta$ formula, Eq. (\ref{bigdel})
for $n_o/\sqrt{N}< 1/2$.  

\section{Distribution of Number of Infections}
We now turn to investigate the distribution of the total number of infections, returning again to the case of a single initial infected person, $n_o=1$.  For the sub- and supercritical cases, these are fairly simple.  In the subcritical case, almost surely the
infection goes extinct before the percentage of infecteds, $I/N$, is significant.  In this case, the up transition probabilities are essentially constant:  $p^+_k = R_0(N-I)/(R_0(N-I) + N)\approx R_0$, $p^-_k = 1- p^+_k \approx 1-R_0$.  Therefore, the random walk reduces to that of a constant leftward bias, with $P_n$ falling exponentially with $n$:
\begin{eqnarray}
P(n) &=&  \frac{R_0^{n-1}}{(1+R_0)^{2n-1}}\left[{{2n-2}\choose{n-1}} - {{2n-2}\choose{n}}\right] \nonumber\\
&\approx& \frac{1}{\sqrt{\pi n^3}}\frac{(4R_0)^{n-1}}{(1+R_0)^{2n-1}} \quad\quad\quad (n \gg 1)
\label{bias}
\end{eqnarray}

\begin{figure}
\center{\includegraphics[width=0.4\textwidth]{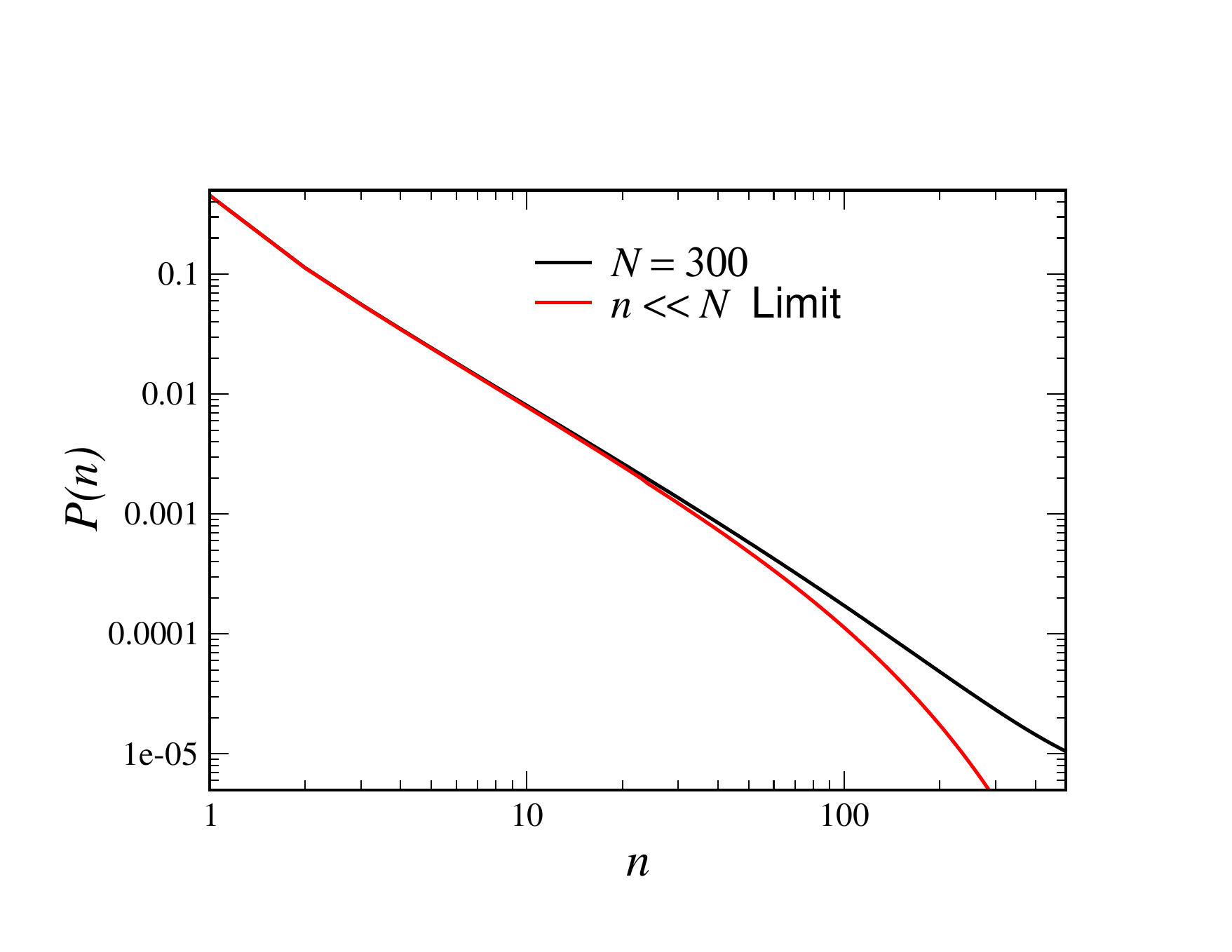}\includegraphics[width=0.4\textwidth]{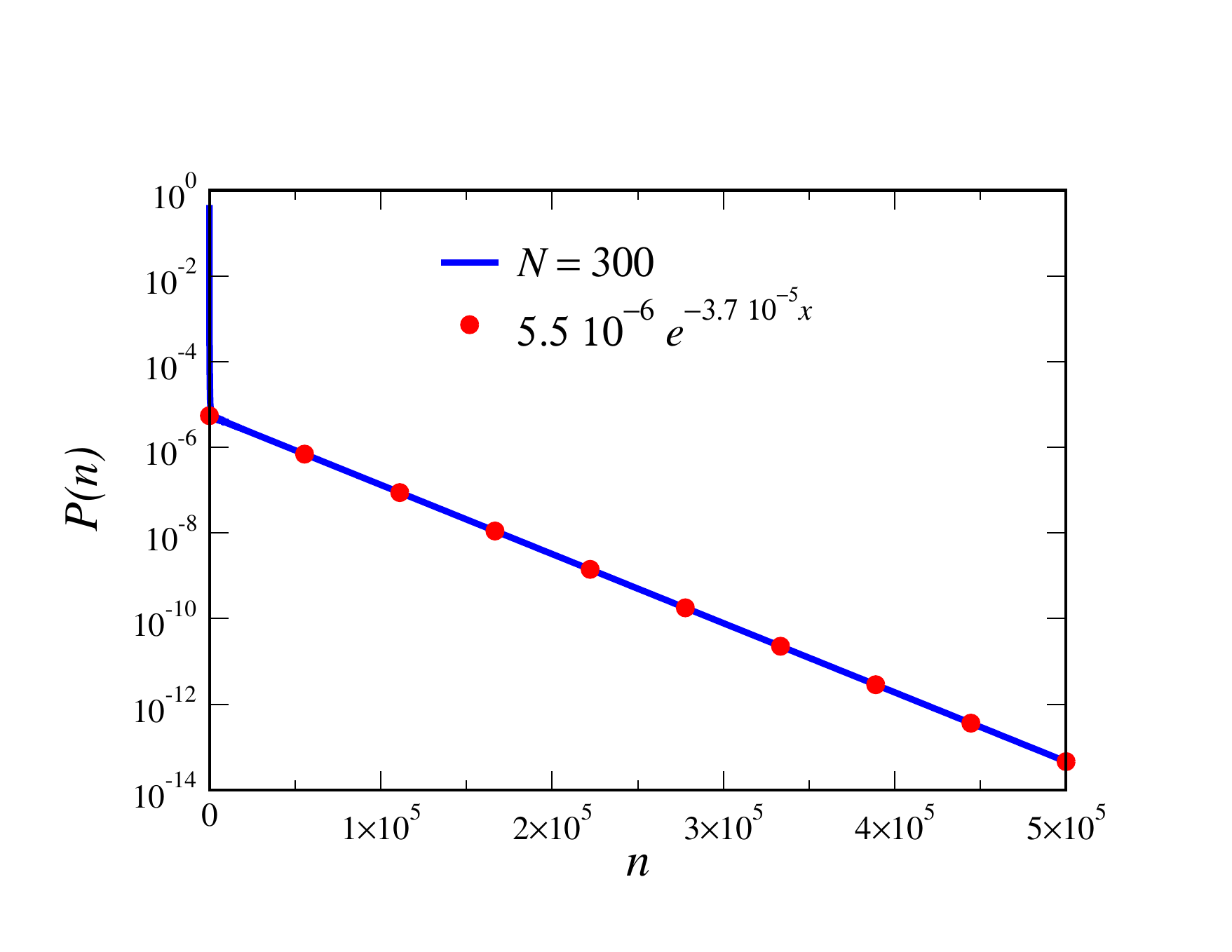}}
\caption{Probability distribution $P(n)$ for total size of epidemic in the supercritical case, $R_0=1.2$, $N=300$.  Left: Behavior for $n < N$, together with the constant bias approximation, Eq. (\ref{bias}). The beginning of the crossover to exponential
behavior is visible toward the end.  Right: Behavior for $n$ on the scale of the average
epidemic size, together with an exponential fit.  The normalization is seen to be $5.5 10^{-6}\cdot 3.7 10^{-5} = 0.2$, consistent with our expectation of $1-1/1.2 = 0.17$.}
\label{dist300_1p2}
\end{figure}

Above threshold, the system spends an exponentially long time in the metastable state.
Thus $P(n)$ for macroscopic $n$'s is an exponential distribution, the waiting time distribution for the decay of the metastable state.
For small epidemics, where the space-dependent drift
is not yet relevant, the system can again be approximated by a random walk with constant bias, this time to the right.  Thus for $n\ll N$,  $P(n)$ is given by Eq. (\ref{bias}).  For larger $N$, $P(n)$ crosses over to a pure exponential decay, normalized to $1-1/R_0$, the probability of the infection surviving to macroscopic size.  This behavior is
exhibited in Fig. \ref{dist300_1p2}.  It is important to note the difference between this
behavior and that exhibited above threshold in the SIR model\cite{Watson,SIR,SIRme}.  There the distribution has
a second peak (in addition to the one at the origin) at the number of infections predicted by the deterministic dynamics.  In the SIS model, the total number of infections above
threshold predicted by the deterministic dynamics is infinite.  Rather the behavior for
major epidemics is the pure exponential waiting-time distribution, with its peak at the origin.

We now turn to investigate the behavior in the crossover regime, $R_0 = 1 + \delta/\sqrt{N}$.  The essential simplification here is in the transition probabilities, which in this
regime can be approximated by
$$
p_\pm = \frac{1}{2} \mp \frac{I}{4N} \pm \frac{\delta}{4\sqrt{N}}
$$
The critical regime is characterized by the scaling $I\sim{\cal{O}}(N)$, so that the
bias is small, of order $1/\sqrt{N}$.  For relatively small Times, ($n \ll \sqrt{N}$) the bias is irrelevant, and the
problem reduces to the unbiased random walk, given by substituting $R_0=1$ in Eq. (\ref{bias}) above:
\begin{eqnarray}
P(n) &=&  \frac{1}{2^{2n-1}}\left[{{2n-2}\choose{n-1}} - {{2n-2}\choose{n}}\right] \nonumber\\
&\approx& \frac{1}{\sqrt{4\pi n^3}}\quad\quad\quad (n \gg 1)
\label{nobias}
\end{eqnarray}
We now study how for larger Times the bias, resulting from the reduction of the susceptible pool with increasing $I$ and the small deviation from criticality, modifies this answer.

\begin{figure}
\center{\includegraphics[width=0.7\textwidth]{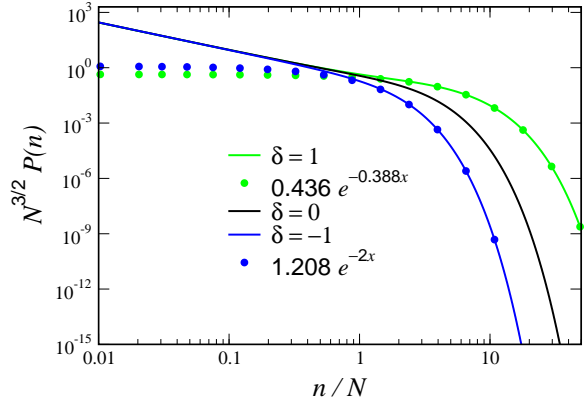}}
\caption{ $N^{3/2}P(n)$, the scaled probability density of epidemics of size $n$ in the critical regime for the cases $\delta=-1$, $0$, and $1$, as a function of scaled epidemic size $n/N$.}
\label{critp}
\end{figure}

As the bias is very weak, however, and only effective at large Times,  we are justified in passing to a Fokker-Planck description for the the probability distribution $K(I,T;I_o)$,
for $I$, given that there were $I_o$ infected individuals at $T=0$.  For a typical major epidemic, $T$ at extinction is of order $N$, this despite the fact that $\bar{n}$ is of order $\sqrt{N}$,
since the probability of a major epidemic is of order $1/\sqrt{N}$.  We thus define $t\equiv T/(2N)$ and $x\equiv I/\sqrt{N}$ and consider $K(x,t;x_o)$:
\begin{equation}
\frac{\partial}{\partial t} K(x,t;x_o)= \frac{\partial^2}{\partial x^2}K + \frac{\partial}{\partial x}\left( x K\right) - \delta \frac{\partial K}{\partial x} 
\label{fp}
\end{equation}
with the initial condition $K(x,t;x_o) = \delta(x-x_o)$ and the absorbing boundary condition $K(0,t;x_o)=0$.  In terms of $K$, the probability distribution for the epidemic size,  $P(n)$, is given by
\begin{equation}
P(n) = N^{-3/2}\left. \frac{\partial^2}{\partial x \partial x_o} K(x,n/N;x_o)\right|_{x=x_o=0}
\label{pk}
\end{equation}
since $x_o = I_o/\sqrt{N} = 1/\sqrt{N}$.  This critical regime distribution function, obtained from a numerical solution of Eqs. (\ref{fp}) and (\ref{pk}), is presented in Fig. \ref{critp} for $\delta=-1$, $0$, and $1$.  We see that for small $n$ all three curves collapse into a universal power law. For large $n$ the distributions fall off rapidly, with the speed of falloff decreasing with increasing $\delta$.

To make more progress, we recognize Eq. (\ref{fp}) as the imaginary-time Schr\"odinger Eq. for a harmonic oscillator potential, modulo a Gaussian similarity transformation. Defining
$$
K(x,t;x_o) = e^{ (x-x_o)\delta/2 - (x^2-x_o^2)/4 +t/2}G(x,t;x_0)
$$
we have
$$
\dot{G} =  G'' - \frac{1}{4}(x-\delta)^2G 
$$
with $G(x,0;x_o)=\delta(x-x_o)$, $G(0,t;x_0)=0$ in terms of which
$$
P(n)=N^{-3/2}e^{n/2N}\left. \frac{\partial^2}{\partial x\partial x_o} G(x,n/N;x_o)\right|_{x=x_o=0}
$$

\begin{figure}
\center{\includegraphics[width=0.7\textwidth]{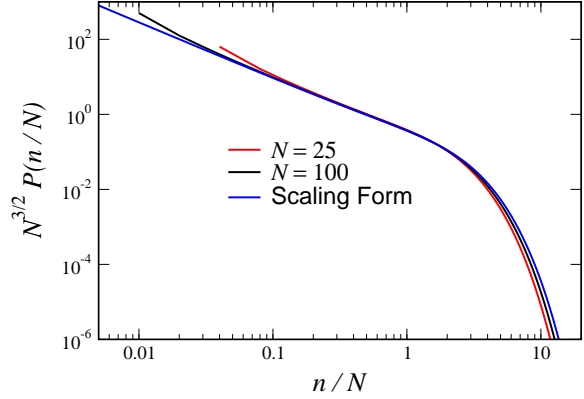}}
\caption{The analytic scaling solution for the threshold $P(n)$, Eq. (\ref{critdisteq}), together with data for $N=25$ and $100$.}
\label{critdist}
\end{figure}

The only complication is the presence of the absorbing boundary condition at $x=0$,
which breaks the reflection symmetry of the potential around $x=\delta$. Exactly at theshold,  $\delta=0$, however, this is not
a problem, as the boundary condition can be enforced by the method of images.  The solution is
$$
G(x,x_o,t) = A(t)\sinh\left(2c(t) x x_m(t)\right)e^{-c(t)(x^2 - x_m(t)^2)}
$$
where
\begin{eqnarray*}
c(t) &=& \frac{1}{4}\coth t\nonumber\\
x_m(t) &=& \frac{x_o}{\cosh t} \nonumber\\
A(t) &=& \frac{1}{\sqrt{\pi}}\sinh^{-1/2} t \, e^{-\frac{x_o^2}{4}\tanh t}
\end{eqnarray*}
This gives the probability distribution 
\begin{equation}
P(n)=  \frac{1}{\sqrt{4\pi N^3}}e^{n/2N} \sinh^{-3/2} (n/N) 
\label{critdisteq}
\end{equation}
This clearly reproduces the expected behavior, Eq. (\ref{nobias}), for $1\ll n \ll N$, and
then decays exponentially for $n \sim N$.  This result is shown in Fig. \ref{critdist}, together with data for $N=25$ and $100$.  Even for these small $N$'s, the agreement is
excellent, exact at the smallest $n$'s, where the discreteness of $n$ factors in.

Knowing $P(n)$ gives us another way to  calculate the mean epidemic size, $\bar{n}$:\begin{eqnarray*}
\bar{n} &=& \int_0^\infty nP(n) = \sqrt{\frac{2N}{\pi}} \int_0^\infty dx \frac{xe^{-x}}{\left(1-e^{-2x}\right)^{3/2}}\nonumber\\
&=& \sqrt{\frac{2N}{\pi}} \int_0^\infty dx \frac{e^{-x}}{\sqrt{1-e^{-2x}}}\nonumber\\
&=&\sqrt{\frac{2N}{\pi}}\left.\sin^{-1}\left(e^{-x}\right)\right|_0^\infty\nonumber\\
&=&\sqrt{\frac{\pi N}{2}}
\end{eqnarray*}
which of course agrees with our previous result, Eq. (\ref{crit}), specialized to $\delta=0$.

We now return to the distribution of $n$ for general $\delta$.  Formally, we can decompose $G(x,t;x_o)$ into a sum over eigenfunctions of the Shroedinger operator:
$$
H = -\frac{d^2}{dx^2} + \frac{1}{4}(x-\delta)^2
$$
as follows:
$$
G(x,x_o,t) = \sum_n \phi_n(x)\phi_n(x_o) e^{-E_n t}
$$
If we define the  Green's function $G(x,x',E)$ as usual by
$$
G(x,x',E) = \sum_n\frac{ \phi_n(x)\phi_n(x')}{E_n-E}
$$
then $G(x,x',-E)$ is 
the Laplace transform of $G(x,x',t)$ with respect to time:
$$
G(x,x',E)=\int_0^\infty dt\, e^{Et} G(x,x',t)
$$
and $G(x,x',E)$ satisfies
$$
HG -EG  = \delta(x-x')
$$
We can recover $G(t)$ from $G(E)$ by an inverse Laplace transform
$$
G(x,x',t)= \frac{1}{2\pi i}\int_{\gamma-i\infty}^{\gamma+i\infty}dE e^{-E t} G(x,x',E)
$$
where $\gamma$ lies to the left of all the poles of $G$. Defining $P(E)$ as:
$$
P(E) \equiv \left.\frac{\partial^2}{\partial x\partial x'} G(x,x',E)\right|_{x=x'=0}
$$
we get an expression for our desired probability distribution $P(n)$:
$$
P(n) = \frac{1}{2\pi i N^{3/2}} \int_{\gamma-i\infty}^{\gamma+i\infty} e^{-(E-1/2) n/N} P(E)
$$

Thus all we need to do is calculated the Green's function $G(x,x',E)$.
There is a nice formula relating
$G(x,x',E)$ to $G^0(x,x',E)$, the Green's function of the problem without the wall.  
The
Green's function for the system with a wall on the left-hand side of the system at $x=a$ is~\cite{Grosche}
$$
G(x,x',E) = G^0(x,x',E) - \frac{G^0(x,a,E)G^0(x',a,E)}{G^0(a,a,E)}
$$
Denoting $f(x,E)$ as the solution of $Hf=Ef$ which decays as $x\to+\infty$, and $g$ the
solution which decays as $x\to-\infty$, then in general
$$
G^0(x,x',E)=\frac{f(x_>)g(x_<)}{\textrm{Wr}[f,g]}
$$
where $\textrm{Wr}[f,g]$ is the Wronskian.
Then, using the fact that the $\textit{Wr}[f,g]$ is constant,
\begin{eqnarray*}
P(E)&\equiv&\left.\frac{\partial^2}{\partial x\partial x'} G(x,x',E)\right|_{x=x'=0}\quad\quad\quad \nonumber\\
&=&\frac{1}{\textrm{Wr}[f,g]} \frac{\partial^2}{\partial x\partial x'} \left[ f(x_>,E)\left(g(x_<,E) - \frac{f(x_<,E)g(0,E)}{f(0,E)}\right)\right]_{x=x'=0}\nonumber\\
&=& \frac{1}{\textrm{Wr}[f,g]}f'(0,E)\left(g'(0,E) - \frac{f'(0,E)g(0,E)}{f(0,E)}\right)\nonumber\\
&=& \frac{f'(0,E)}{f(0,E)}\nonumber\\
\end{eqnarray*}
In our case, $f(x,E)=U(-E,x-\delta)$ where $U$ is that parabolic cylinder function~\cite{AbS} which decays for positive argument.  Thus, 
$$
P(E)=\frac{U'(-E,-\delta)}{U(-E,-\delta)}
$$

The Green's function is also of course the moment generating function.  Thus,
already at this stage, we can use the Green's function to recover the mean epidemic size.
  The answer is
\begin{eqnarray*}
\bar{n} &=&\int_0^\infty nP(n) dn\nonumber\\
&=&  N^{-3/2}\frac{\partial^2}{\partial x\partial x'} \left[\int_0^\infty dn\, e^{n/2N} n K(x,x',n/N)\right]_{x=x'=0}\nonumber\\
&=&N^{1/2} \left. \frac{\partial^2}{\partial x\partial x'} \frac{\partial}{\partial E}G(x,x',E)\right|_{x=x'=0;E=1/2}\nonumber\\
&=&N^{1/2}\left.\frac{\partial}{\partial E} P(E)\right|_{E=1/2}\nonumber\\
&=&-N^{1/2}\left.\frac{\partial^2 }{\partial a\partial x} \ln(U(a,x))\right|_{a=-1/2,x=-\delta}
\end{eqnarray*}
We thus need an expression for $U(a,x)$ for $a$ near $-1/2$.  We can get this by
perturbing about the Gaussian solution at $a=-1/2$, writing
$$
\psi= Ae^{-x^2/4} + \psi_1
$$
where $\psi_1$ satisfies the inhomogeneous equation
$$
-\psi_1'' + \frac{1}{4}\psi_1 - \frac{1}{2}\psi_1 = \delta_a A e^{-x^2/4}
$$
where $\delta_a$ is the shift in $a$.  Then, since the two modes of the
homogenous equation are $f_1=e^{-x^2/4}$ and $f_2=e^{-x^2/4}\int_0^x dt e^{t^2/2}$,
the solution for $\psi_1$ that decays as $x \to +\infty$  is:
$$
\psi_1 =-\delta_a f_1(x)\int_0^x dx' Ae^{-x'^2/4} f_2(x') - \delta_a f_2(x) \int_x^\infty dx'
A e^{-x'^2/2}
$$
The derivative w.r.t. $a$ is then
$$
\left.\frac{\partial}{\partial a} \ln(U(a,x))\right|_{a=-1/2} = \frac{\psi_1}{\delta_a \psi_0} 
= -\int_0^x dx' e^{-x'^2/4} f_2(x') -  f_2(x) e^{x^2/4} \int_x^\infty dx'  e^{-x'^2/2}
$$
We can now differentiate w.r.t. $x$ and get
\begin{eqnarray*}
\bar{n} &=& -N^{1/2}\left[-e^{-x^2/4} f_2(x) - e^{x^2/2}\int_x^\infty dx' e^{-x'^2/2} + f_2(x) e^{-x'^2/4}\right]_{x=-\delta}\nonumber\\
&=& \sqrt{\frac{\pi N}{2}} e^{\delta^2/2} \left[ 1 + \textrm{erf}\left(\frac{\delta\sqrt{2}}{2}\right) \right]
\end{eqnarray*}
reproducing of course our previous result, Eq. (\ref{crit})!

The last order of business is to take the inverse Laplace transform.  We have
$$
P(n)=\frac{1}{2\pi i N^{3/2}} e^{n/2N}\int_{\gamma-i\infty}^{\gamma+i\infty} e^{an/N} \frac{U'(a,-\delta)}{U(a,-\delta)}
$$
For the threshold case, $\delta=0$, we have 
$$
\frac{U'(a,0)}{U(a,0)}=-\sqrt{2}\frac{\Gamma\left(\frac{3}{4}+\frac{a}{2}\right)}{\Gamma\left(\frac{1}{4}+\frac{a}{2}\right)}
$$
The integral can then be done by residues, giving
$$
P(n)=\sqrt{\frac{2}{\pi N^3}} \sum_{k=0}^\infty e^{-(2k+1)n}\frac{(2k+1)!!}{k!2^k}=\sqrt{\frac{2}{\pi}}\frac{e^{-n}}{(1-e^{-2n})^{3/2}}
$$
in agreement with our previous result.  For general $\delta$, however, one has to do the integral numerically.  The most important information however, the asymptotic behavior for small and large $n$, can be gleaned analytically.

We can  get the small $n$ expansion of the
distribution function by using the large $a$ expansion of the integrand.  Using the Hankel Countour Integral:
$$
\frac{1}{2\pi i} \int_{\cal{C}} t^{-z}e^{t} dt = \frac{1}{\Gamma(z)}
$$
 we have
\begin{eqnarray*}
P(n)&\approx& e^{n/2N}\frac{1}{2\pi i N^{3/2}}\int_{\gamma-i\infty}^{\gamma+i\infty} da e^{an}\left[
-\sqrt{a}-\frac{\delta^2}{8}a^{-1/2} + \frac{\delta}{8}a^{-1} + \left(-\frac{1}{16} + \frac{\delta^4}{128}\right)a^{-3/2} + \ldots\right]\nonumber\\
&\approx& N^{-3/2}e^{n/2N} \left[ \frac{-1}{\Gamma(-1/2)}\left(\frac{n}{N}\right)^{-3/2} - \frac{\delta^2}{8\Gamma(1/2)}\left(\frac{n}{N}\right)^{-1/2} + \frac{\delta}{8\Gamma(1)}\left(\frac{n}{N}\right)^{0}  + \left(-\frac{1}{16} + \frac{\delta^4}{128}\right)\frac{1}{\Gamma(3/2)}\left(\frac{n}{N}\right)^{1/2} + \ldots\right] \nonumber\\
&\approx& \frac{e^{n/2N} }{\sqrt{4\pi n^3}} \left[1 - \frac{\delta^2}{4}\frac{n}{N} + \frac{\delta\sqrt{\pi}}{4}\left(\frac{n}{N}\right)^{3/2} -
\frac{1}{4}\left(1 - \frac{\delta^4}{8}\right)\left(\frac{n}{N}\right)^2  + \ldots\right] 
\end{eqnarray*}
Thus we have that the leading order behavior at small $n$ is $P(n)\approx (4\pi n^3)^{-1/2}$ independent of $\delta$. One can also verify that this small $n$ series reproduces the
full $P(n)$ for the $\delta=0$ case.

The large-$n$ asymptotics is clearly given by the ground state of the wall problem.  In the limit $\delta\to \infty$, the wall, relative to the bottom of the potential, moves to $-\infty$ and the ground state
energy goes to $1/2$ in our units.  This translates to a decay rate of zero, once the 
$exp(n/2N)$ factor is taken into account.  As $\delta$ decreases the wall moves closer
to the potential minimum and the energy (and the decay rate) rise monotonically.  The
energy is $3/2$ in our units when the wall hits the potential minimum at $\delta=0$.
This leads to the decay behavior $exp(-n/N)$ at threshold, in accord with the full solution in this case.  The energy continues to rise as $\delta$ is decreased below 0, leading to
a faster decay in $n$, with the energy diverging in the limit $\delta \to -\infty$.

For example,  the second excited state of the harmonic oscillator has  zeroes
located a distance $1/\sqrt{2\omega}=1$ to the right and left of the energy minimum. Thus, for $\delta=-1$, we expect the decay $exp(-2n/N)$.  Since the normalized eigenvector in our units
is
$$
\phi(x) = {\cal{N}}\left[x^2-1\right]e^{-x^2/4}
$$
where ${\cal{N}}^{-2} = 2e^{-1/2}+\sqrt{2\pi}\textrm{erfc}(\sqrt{2}/2)$
so that $(\phi'(1))^2 = 4{\cal{N}}^2e^{-1/2}\approx 1.208$, then for large $n$,
$$
P(n)\sim 1.208 e^{-2n/N}
$$
For $\delta=+1$ on the other hand, the solution of the Schroedinger problem is given
by the first zero of $U(a,-1)$ at $a_1=-0.88824=-E_1$, leading to a decay exponent of $0.388824$. The coefficient of the exponent is given by $U'(a_1,-1)/(\frac{\partial}{\partial a} U(a_1,-1)) = 0.4365$. 
This  can be seen in Fig. \ref{critp}, where the correct exponential falloff in both cases is
seen for large $n$.  For $n/N<2$, the effect of the higher eigenvectors leads to deviation
from a pure exponential behavior.  In general,
the large $n$ approximation is accurate as long as the next higher eigenvector has
decayed.  This "energy gap" is approximately $1$ for large positive $\delta$, rises to
2 at $\delta=0$ and continues to rise as $\delta$ becomes more negative.  However
as the ratio of the energy gap to the ground state energy fails as $\delta$ decreases,
the role of the higher excited states becomes more pronounced as $\delta$ decreases.

Examining the limit of large positive $\delta$ in more detail,  the energy is very slightly
above $1/2$, so there is an extremely small decay rate.  The actual ground state
energy can be calculated as follows:  We first shift $x$ by $\delta$, so that the quadratic potential is centered at the origin. Then, since the ground state energy in the presence of the wall is close to the wall-free value of $1/2$, we can write
$$
\psi= Ae^{-x^2/4} + \psi_1
$$
where $\psi_1$ satisfies the inhomogeneous equation
$$
-\psi_1'' + \frac{1}{4}\psi_1 - \frac{1}{2}\psi_1 = \epsilon A e^{-x^2/4}
$$
and $\epsilon$ is the shift in the energy.  As before, the two modes of the 
homogenous equation are $f_1=e^{-x^2/4}$ and $f_2=e^{-x^2/4}\int_0^x dt e^{t^2/2}$,
The solution for $\psi_1$ that decays as $x \to +\infty$  is:
$$
\psi_1 =-\epsilon f_1(x)\int_0^x dx' Ae^{-x'^2/4} f_2(x') - \epsilon f_2(x) \int_x^\infty dx'
A e^{-x'^2/2}
$$
The second term dominates for large negative $x$, so that
$$
\psi(x)\approx Ae^{-x^2/4} + \epsilon A \frac{\sqrt{2\pi}}{x}e^{x^2/4}
$$
 and then the boundary condition that $\psi(-\delta)=0$ gives
$$
\epsilon \approx \frac{\delta}{\sqrt{2\pi}}e^{-\delta^2/2}
$$
Then,
$$
\frac{d\psi}{d a} = - \frac{d\psi_1}{d\epsilon} = \frac{\sqrt{2\pi}}{\delta}Ae^{\delta^2/4}
$$
and
$$
\psi'(-\delta) = A\left[\frac{\delta}{2}e^{-\delta^2/4} + \epsilon \sqrt{\frac{\pi}{2}}e^{\delta^2/4}\right] = A\delta e^{-\delta^2/4}
$$
This gives for the leading order asymptotics of $P(n)$ for large $n$, large $\delta$:
$$
P(n) \approx \frac{\delta^2}{\sqrt{2\pi N^3}}e^{-\delta^2/2}\exp\left(\frac{n}{N\sqrt{2\pi}}\delta e^{-\delta^2/2}\right)
$$
so that the total probability of a major epidemic is the integral of $P(n)$, which is $\delta/\sqrt{N}\approx 1-1/R_0$, as expected, and
$$
\bar{n} \approx \sqrt{2\pi}e^{\delta^2/2}
$$
in accord with our previous result.

In the limit of large $n$ and large negative $\delta$, we get that the ground state
energy is large, approximately $E\approx\delta^2/4$, so that $a\approx -\delta^2/4$.  However, there are many states with approximately this energy.  It is best to work
directly from our integral formulation.  For large negative $\delta$, we can use a WKB
ansatz to write $U(a,x) \approx \exp(\sqrt{-a}S(x/\sqrt{-a})$ which yields
$$
\frac{U'(a,x)}{U(a,x)} = S' = -\sqrt{a + x^2/4}
$$
Expanding this in a power series for large $a$, we get
$$
S'(-\delta) = -\sqrt{a} \sum_{k=0}^\infty \frac{\Gamma(3/2)}{k!\Gamma(3/2-k)}\left(\frac{\delta^2}{4a}\right)^k
$$
Performing the integral over $a$ gives
\begin{eqnarray*}
P(n)&=& -N^{-3/2}e^{n/2N} \sum_{k=0}^\infty \frac{\Gamma(3/2)}{k!\Gamma(3/2-k)\Gamma(k-1/2)}\left(\frac{\delta^2}{4}\right)^k \left(\frac{n}{N}\right)^{k-3/2}\nonumber\\
&=&- e^{n/2N} n^{-3/2} \sum_{k=0}^\infty \frac{\Gamma(3/2)\sin(\pi(k-1/2))}{k! \pi }\left(\frac{\delta^2 n}{4N}\right)^k \nonumber\\
&=& \frac{e^{n/2N}}{\sqrt{4\pi n^3}} e^{-n\delta^2/4N}
\end{eqnarray*}
We see that we have successfully summed all the leading order contributions.  For
large negative $\delta$, this is cut off at $n$'s of order $1/\delta^2$, so we may drop the
$e^{n/2N}$ term.  Then, we have the result for a pure constant drift, and so matches on
to the subcritical distribution (which is the same as the SIR case).  It of course reproduces the correct mean as well, since in the integral over $n$,
 small $n$'s predominate, and
$$
\bar{n}\approx \int_0^\infty  dn\, n\, e^{-n\delta^2/4N} \frac{1}{2\sqrt{\pi}n^{3/2}} = \frac{\sqrt{N}}{(-\delta)}
$$
This in turn matches on to the subcritical result, $\bar{n}\approx 1/(1-{R_0})$, as ${R_0}$
approaches one from below.  

An interesting subtlety arise if we consider the zeroth moment of the distribution.  In normal circumstances this would be unity, but the scaling behavior of $P(n)$ dictates
that the normalization integral is formally of order $N^{-1/2}$ and furthermore diverges,
due to the $n^{-3/2}$ behavior of $P$ for small $n$. Nevertheless, if we blindly forge ahead, we find
\begin{eqnarray*}
P_{\textit{tot}} &=& \int_0^\infty dn P(n) \nonumber\\
&=& N^{-1/2} P(1/2)\nonumber\\
&=& N^{-1/2}\frac{U'(-1/2,-\delta)}{U(-1/2,-\delta)}\nonumber\\
&=& \frac{\delta}{2\sqrt{N}}
\end{eqnarray*}
Clearly this finite answer is the result of an analytic continuation.  To understand its
significance,  let us consider the
difference between $P(n)$ and the distribution for the constant bias random walk with
the same $\delta$.  For small $n$ these distributions as we have seen are identical,
so the difference is integrable.  Integrating the constant bias random walk, we get
$$
P_{\textit{tot}}^{CB} = \int_0^\infty dn \frac{1}{\sqrt{4\pi n^3}} e^{-n\delta^2/4N} = -\frac{|\delta|}{2\sqrt{N}}
$$
Thus, the difference is  $(\delta+|\delta|)/(2\sqrt{N})$.  This is zero for $\delta \le 0$,
which is correct, since even without the space-dependent drift toward the origin, every
walker will eventually hit the origin.  On the other hand, for $\delta>0$ the difference is $\delta/\sqrt{N}$, which reflects the fact that with the added space-dependent drift all walkers are
guaranteed to return to the origin, while without only a fraction $1/R_0 \approx 1-\delta/\sqrt{N}$ do.  Thus, looking at the difference between distributions provides an excellent way to make rigorous the concept of "major epidemics", even slightly above threshold in the critical regime. Even below threshold, it highlights the added role of the space-dependent drift in reducing the probability of larger epidemics in favor of smaller ones. 

\section{Concluding Discussion}
We have exhibited an exact expression for the mean epidemic size in the SIS model of endemic infection.  We have evaluated this in the limit of large population size, and shown the crossover behavior that occurs in the vicinity of the critical infection number,
$R_0=1$.  We have also calculated the distribution function for the epidemic size, again focussing on the crossover regime.  

It is important to note that the crossover behavior is universal, independent of the details of the model.  What is important is the existence of two fixed points of the rate equation dynamics and a critical parameter where the two fixed points interchange stability.  In the case the functional form of the mean epidemic size as a function of $\delta$, the scaled distance to the critical point will be the same, along with the scaling behavior with $N$.  This is also true in general for any first passage time problem where the transition rates are constant (independent of $N$ and location) at the transition.  The first passage time,
i.e. the mean (physical) time to extinction, in the SIS model also exhibits a crossover behavior at the transition, albeit different than that of the mean epidemic size.  This is due to the fact that the transition rates in time are location dependent, $p^+_k = \alpha k (N-k)/N$, $p^-_k = \beta k$.  As Doering, et al.~\cite{SanDoer} did not investigate the transition behavior of the mean extinction time, we for completeness present it here.  The mean extinction time (starting with one infected) in the crossover regime is given by
\begin{eqnarray*}
\bar{t}(\delta) &\approx& \sum_{k=1}^\infty \frac{1}{\beta k} e^{-k^2/2N + \delta k/\sqrt{N}} \nonumber\\
&\approx& \gamma + \frac{1}{2}\ln N + \int_0^\infty  \ln x e^{-x^2/2 + \delta x} (x - \delta) dx 
\end{eqnarray*}
In particular, at threshold, the integral can be performed analytically at we get 
$$
\bar{t}(\delta=0) \approx \frac{1}{2}\left(\ln 2N + \gamma\right)
$$

The SIR model, as has been demonstrated~\cite{SIR,BnK,SIRme}, exhibits a different scaling in the threshold regime since its fixed point structure is different.  The SIR model has a line of fixed points at $I=0$, but no fixed point at finite $I$.  In fact, any tendency to immunity (or death, for that matter) will cause an otherwise SIS model to exhibit SIR behavior in the threshold regime for large enough $N$.  This is due to the fact that the
Time dependent bias in the SIR model, no matter how small in strength, overwhelms the space-dependent bias~\cite{SIRme} for large enough $N$.

\acks
The author acknowledges the support of the Israel Science Foundation.  He thanks
I. N\.asell for suggesting the problem and N. Shnerb for discussions. 
\bibliographystyle{apt}
\bibliography{sis1}
\end{document}